\documentclass[reqno,11pt]{amsart}
\usepackage{graphicx}
\usepackage{amscd}
\usepackage{slashed}
\usepackage{amssymb}
\usepackage{ushort}
\usepackage{mathtools} 
\usepackage{epsfig}
\usepackage{pst-grad} 
\usepackage{pst-plot} 
\usepackage[mathscr]{eucal}
\numberwithin{equation}{section}
\allowdisplaybreaks[1]

\title[The Chiral Index of the Fermionic Signature Operator]
{The Chiral Index of the Fermionic Signature Operator}

\author[F.\ Finster]{Felix Finster \\ \\ April 2014}
\address{Fakult\"at f\"ur Mathematik \\ Universit\"at Regensburg \\ D-93040 Regensburg \\ Germany}
\email{finster@ur.de}
\thanks{Supported in part by the Deutsche Forschungsgemeinschaft.}

\newtheorem{Def}{Definition}[section]
\newtheorem{Thm}[Def]{Theorem}
\newtheorem{Prp}[Def]{Proposition}

\newtheorem{Example}[Def]{Example}

\newcommand{\Thanks}{\vspace*{.5em} \noindent \thanks}
\newcommand{\beq}{\begin{equation}}
\newcommand{\eeq}{\end{equation}}
\newcommand{\Proof}{\begin{proof}}
\newcommand{\QED}{\end{proof} \noindent}
\newcommand{\QEDrem}{\ \hfill $\Diamond$}
\newcommand{\la}{\langle}
\newcommand{\ra}{\rangle}
\newcommand{\bra}{\mathopen{<}}
\newcommand{\ket}{\mathclose{>}}
\newcommand{\Sl}{\mathopen{\prec}}
\newcommand{\Sr}{\mathclose{\succ}}

\newcommand{\C}{\mathbb{C}}
\newcommand{\R}{\mathbb{R}}
\newcommand{\Q}{\mathbb{Q}}
\newcommand{\1}{\mbox{\rm 1 \hspace{-1.05 em} 1}}
\newcommand{\Z}{\mathbb{Z}}
\newcommand{\N}{\mathbb{N}}

\newcommand{\nuslsh}{\slashed{\nu}}
\renewcommand{\H}{\mathscr{H}}

\newcommand{\U}{{\rm{U}}}

\newcommand{\e}{{\mathfrak{e}}}

\newcommand{\uslsh}{\slashed{u}}

\newcommand{\bep}{\begin{pmatrix}}
\newcommand{\enp}{\end{pmatrix}}

\renewcommand{\O}{\mathscr{O}}

\newcommand{\Dir}{{\mathcal{D}}}
\newcommand{\D}{{\mathscr{D}}}

\newcommand{\B}{{\mathscr{B}}}
\renewcommand{\O}{{\mathscr{O}}}

\newcommand{\Lin}{\text{\rm{L}}}

\newcommand{\Cisc}{C^\infty_{\text{sc}}}

\newcommand{\Span}{\text{span}}

\DeclareMathOperator{\supp}{supp}

\DeclareMathOperator{\ind}{ind}

\newcommand{\F}{{\mathscr{F}}}
\newcommand{\Sig}{\mathscr{S}}
\newcommand{\pseudo}{\Gamma}
\newcommand{\scrM}{\mycal M}
\newcommand{\scrN}{\mycal N}

\DeclareFontFamily{OT1}{rsfso}{}
\DeclareFontShape{OT1}{rsfso}{m}{n}{ <-7> rsfso5 <7-10> rsfso7 <10-> rsfso10}{}
\DeclareMathAlphabet{\mycal}{OT1}{rsfso}{m}{n}

\begin{document}
\maketitle

\begin{abstract}
We define an index of the fermionic signature operator 
on even-dimen\-sio\-nal globally hyperbolic spin manifolds of finite lifetime.
The invariance of the index under homotopies is studied.
The definition is generalized to causal fermion systems with a chiral grading.
We give examples of space-times and Dirac operators thereon for which
our index is non-trivial.
\end{abstract}

\tableofcontents

\section{Introduction}
In the recent papers~\cite{finite, infinite} the fermionic signature operator was introduced
on globally hyperbolic Lorentzian spin manifolds. It is a bounded symmetric operator on the Hilbert space
of solutions of the Dirac equation which depends on the global geometry of space-time.
This raises the question how the geometry of space-time is related to spectral properties of
the fermionic signature operator. The first step in developing the resulting ``Lorentzian spectral geometry''
is the paper~\cite{drum} where the simplest situation of  Lorentzian surfaces is considered.
In the present paper, we proceed in a somewhat different direction and 
show that there is a nontrivial index associated to the fermionic signature operator.
This is the first time that an index is defined for a geometric operator on a Lorentzian manifold.

We make essential use of the decomposition of spinors in even space-time dimension
into left- and right-handed components (the ``chiral grading''). The basic idea is to
decompose the fermionic signature operator~$\Sig$ using the chiral grading as
\beq \label{Sigprop}
\Sig = \Sig_L + \Sig_R \qquad \text{with} \qquad \Sig_L^* = \Sig_R \:,
\eeq
and to define the so-called {\em{chiral index}} of~$\Sig$ as the Noether index of~$\Sig_L$.
After providing the necessary preliminaries (Section~\ref{secprelim}), this definition will be given in
Section~\ref{secindex} in space-times of finite lifetime.
In order to work out the mathematical essence of our index, in Section~\ref{seccfs} we 
also give its definition in the general setting of causal fermion systems
(for an introduction to causal fermion systems see~\cite{rrev} or~\cite{topology}).
Section~\ref{secodd} is devoted to a variant of the chiral index which applies in the special case
of the massless Dirac equation and a Dirac operator which is odd with respect to the chiral grading.
In Section~\ref{secstable} we analyze the invariance properties of the chiral indices when
space-time or the Dirac operator are deformed by a homotopy.
In Sections~\ref{secex1}--\ref{secex3} we construct examples of fermionic signature operators
with a non-trivial index and illustrate the homotopy invariance.
Finally, in Section~\ref{secoutlook} we discuss our results and and give an outlook on
potential extensions and applications, like the generalization to space-times of infinite lifetime.

We point out that the purpose of this paper is to define the chiral index, to study a few basic
properties and to show in simple examples that it is in general non-trivial.
But we do not work out any physical applications, nor do we make the connection
to geometric or topological invariants.
These shortcomings are mainly due to the fact that we only succeeded in computing
the index explicitly in highly symmetric and rather artificial examples.
Moreover, it does not seem easy to verify the conditions needed for the homotopy invariance.
For these reasons, we leave physically interesting examples and geometric stability results
as a subject of future research.
All we can say for the moment is that the chiral index describes the ``chiral asymmetry''
of the Dirac operator in terms of an integer. This integer seems to depend on the geometry of the
boundary of space-time and on the singular behavior of the potentials in the
Dirac equation. Smooth potentials in the Dirac equation, however, tend to not affect the
index.

\section{Preliminaries} \label{secprelim}
We recall a few basic constructions from~\cite{finite}. Let~$(\scrM, g)$ be a smooth, globally
hyperbolic Lorentzian spin manifold of even dimension~$k \geq 2$.
For the signature of the metric we use the convention~$(+ ,-, \ldots, -)$.
We denote the spinor bundle by~$S\scrM$. Its fibers~$S_x\scrM$ are endowed
with an inner product~$\Sl .|. \Sr_x$ of signature~$(n,n)$
with~$n=2^{k/2-1}$ (for details see~\cite{baum, lawson+michelsohn}),
which we refer to as the spin scalar product. Clifford multiplication is described by a mapping~$\gamma$
which satisfies the anti-commutation relations,
\[ \gamma \::\: T_x\scrM \rightarrow \Lin(S_x\scrM) \qquad
\text{with} \qquad \gamma(u) \,\gamma(v) + \gamma(v) \,\gamma(u) = 2 \, g(u,v)\,\1_{S_x(\scrM)} \:. \]
We write Clifford multiplication in components with the Dirac matrices~$\gamma^j$
and use the short notation with the Feynman dagger, $\gamma(u) \equiv u^j \gamma_j \equiv \uslsh$.
The metric connections on the tangent bundle and the spinor bundle are denoted by~$\nabla$.

In the even-dimensional situation under consideration, the spinor bundle has a decomposition
into left- and right-handed components. We describe this chiral grading by an operator~$\pseudo$
(the ``pseudoscalar operator,'' in physics usually denoted by~$\gamma^5$),
\[ \pseudo \::\: S_x\scrM \rightarrow S_x\scrM \:, \]
having for all~$u \in T_x\scrM$ the properties
\beq \label{gammaprop}
\pseudo^* = -\pseudo\:,\qquad \pseudo^2 = \1 \:,\qquad 
\pseudo\, \gamma(u) = -\gamma(u) \,\pseudo \:,\qquad \nabla \pseudo = 0
\eeq
(where the star denotes the adjoint with respect to the spin scalar product).
We denote the chiral projections to the left- and right-handed components by
\beq \label{chiLRdef}
\chi_L = \frac{1}{2} \big( \1 - \pseudo \big) \qquad \text{and} \qquad \chi_R = \frac{1}{2} \big( \1 + \pseudo \big) \:.
\eeq

The sections of the spinor bundle are also referred to as wave functions.
We denote the smooth sections of the spinor bundle by~$C^\infty(\scrM, S\scrM)$.
Similarly, $C^\infty_0(\scrM, S\scrM)$ denotes the smooth sections with compact support.
On the compactly supported wave functions, one can introduce the Lorentz invariant
inner product
\begin{gather}
\bra .|. \ket \::\: C^\infty_0(\scrM, S\scrM) \times C^\infty_0(\scrM, S\scrM) \rightarrow \C \:, \\
\bra \psi|\phi \ket := \int_\scrM \Sl \psi | \phi \Sr_x\: d\mu_\scrM\:. \label{stip}
\end{gather}
The Dirac operator~$\Dir$ is defined by
\[ \Dir := i \gamma^j \nabla_j + \B \::\: C^\infty(\scrM, S\scrM) \rightarrow C^\infty(\scrM, S\scrM)\:, \]
where~$\B \in \Lin(S_x)$ (the ``external potential'') typically is a smooth multiplication operator
which is symmetric with respect to the spin scalar product.
In some of our examples, $\B$ will be chosen more generally as a convolution operator
which is symmetric with respect to the inner product~\eqref{stip}.
For a given real parameter~$m \in \R$ (the ``mass''), the Dirac equation reads
\[ (\Dir - m) \,\psi = 0 \:. \]
We mainly consider solutions in the class~$\Cisc(\scrM, S\scrM)$ of smooth sections
with spatially compact support. On such solutions one has the scalar product
\beq \label{print}
(\psi | \phi) = 2 \pi \int_\scrN \Sl \psi | \nuslsh \phi \Sr_x\: d\mu_\scrN(x) \:,
\eeq
where~$\scrN$ denotes any Cauchy surface and~$\nu$ its future-directed normal.
Due to current conservation, the scalar product is
independent of the choice of~$\scrN$ (for details see~\cite[Section~2]{finite}).
Forming the completion gives the Hilbert space~$(\H_m, (.|.))$.

For the construction of the fermionic signature operator, we need to
extend the bilinear form~\eqref{stip} to the solution space of the Dirac equation.
In order to ensure that the integral in~\eqref{stip} exists, we need
to make the following assumption (for more details see~\cite[Section~3.2]{finite}).
\begin{Def} \label{defmfinite}
A globally hyperbolic space-time~$(\scrM,g)$ is said to be {\bf{{\em{m}}-finite}} if
there is a constant~$c>0$ such that for
all~$\phi, \psi \in \H_m \cap \Cisc(\scrM, S\scrM)$, the
function~$\Sl \phi | \psi \Sr_x$  is integrable on~$\scrM$ and
\[ |\bra \phi | \psi \ket| \leq c \:\|\phi\|\: \|\psi\| \]
(where~$\| . \| = (.|.)^\frac{1}{2}$ is the norm on~$\H_m$).
\end{Def} \noindent
Under this assumption, the space-time inner product is well-defined as
a bounded bilinear form on~$\H_m$,
\[ \bra .|. \ket \::\: \H_m \times \H_m \rightarrow \C\:. \]
Applying the Riesz representation theorem, we can
uniquely represent this bilinear form with a signature operator~$\Sig$,
\beq \label{Sdef}
\Sig \::\: \H_m \rightarrow \H_m \qquad \text{with} \qquad
\bra \phi | \psi \ket = ( \phi \,|\, \Sig\, \psi) \:.
\eeq
We refer to~$\Sig$ as the {\bf{fermionic signature operator}}.
It is obviously a bounded symmetric operator on~$\H_m$.
We note that the construction of the fermionic signature operator is manifestly covariant
and independent of the choice of a Cauchy surface.

\section{The Chiral Index} \label{secindex}
We now modify the construction of the fermionic signature operator by inserting the chiral projection
operators into~\eqref{stip}. We thus obtain the bilinear forms
\beq \label{stipLR}
\bra \psi|\phi \ket_{L\!/\!R} = \int_\scrM \Sl \psi \,|\, \chi_{L\!/\!R} \,\phi \Sr_x\: d\mu_\scrM\:.
\eeq
For the space-time integrals to exist, we need the following assumption.
\begin{Def} \label{defGfinite}
A globally hyperbolic space-time~$(\scrM,g)$ is said to be {\bf{$\pseudo$-finite}} if
there is a constant~$c>0$ such that for
all~$\phi, \psi \in \H_m \cap \Cisc(\scrM, S\scrM)$, the
function~$\Sl \phi | \pseudo \psi \Sr_x$  is integrable on~$\scrM$ and
\[ |\bra \phi | \pseudo \psi \ket| \leq c \:\|\phi\|\: \|\psi\| \:. \]
\end{Def} \noindent
There seems no simple relation between $m$-finiteness and $\Gamma$-finiteness.
But both conditions are satisfied if we assume that the space-time~$(\scrM,g)$ has {\bf{finite lifetime}}
in the sense that it admits a foliation~$(\scrN_t)_{t \in (t_0, t_1)}$ by Cauchy surfaces with~$t_0, t_1 \in \R$
such that the function~$\la \nu, \partial_t \ra$ is bounded on~$\scrM$
(see~\cite[Definition~3.4]{finite}).
The following proposition is an immediate generalization of~\cite[Proposition~3.5]{finite}.
\begin{Prp} Every globally hyperbolic manifold of finite lifetime is $m$-finite
and $\Gamma$-finite.
\end{Prp}
\Proof Let~$\psi \in \H_m \cap \Cisc(\scrM, S\scrM)$ and~$C(x)$ one of the operators~$\1_{S_x}$ or~$i \pseudo(x)$.
Applying Fubini's theorem and decomposing the volume measure, we obtain
\[ \bra \psi | C \psi \ket = \int_\scrM \Sl \psi | C \psi \Sr(x)\: d\mu_\scrM(x) \\
=\int_{t_0}^{t_1} \int_{\scrN_t} \Sl \psi | C \psi \Sr\, \la \nu, \partial_t \ra \,dt \,d\mu_{\scrN_t} \]
and thus
\[ \big| \bra \psi | C \psi \ket \big| \leq \sup_\scrM \la \nu, \partial_t \ra
\int_{t_0}^{t_1} dt \int_{\scrN_t} |\Sl \psi | C \psi \Sr|\,d\mu_{\scrN_t} \:. \]
Rewriting the integrand as
\[ |\Sl \psi | C \psi \Sr| = |\Sl \psi | \nuslsh\, (\nuslsh C) \psi \Sr| \:, \]
the bilinear form~$\Sl .| \nuslsh . \Sr$ is a scalar product. Moreover, the operator~$\nuslsh C$
is symmetric with respect to this scalar product. Using that
\[ (\nuslsh)^2 = \1 = (i \nuslsh \pseudo)^2 \:, \]
we conclude that the sup-norm corresponding to the
scalar product~$\Sl .| \nuslsh . \Sr$ of the operator~$\nuslsh C$ is equal to one. Hence
\[ \int_{\scrN_t} |\Sl \psi | C \psi \Sr|\,d\mu_{\scrN_t} \leq
\int_{\scrN_t} \Sl \psi | \nuslsh \psi \Sr\,d\mu_{\scrN_t} = (\psi | \psi) \:, \]
and consequently
\[ \big| \bra \psi | C \psi \ket \big| \leq (t_1-t_0)\: \sup_\scrM \la \nu, \partial_t \ra\; \|\psi\|^2\:. \]
Polarization and a denseness argument give the result.
\QED

Assuming that our space-time is $m$-finite and $\pseudo$-finite, the bilinear forms~\eqref{stipLR}
are bounded on~$\H_m \times \H_m$. Thus we may represent them with respect to the Hilbert space scalar
product in terms of signature operators~$\Sig_{L\!/\!R}$,
\beq \label{SLRdef2}
\Sig_{L\!/\!R} \::\: \H_m \rightarrow \H_m \qquad \text{with} \qquad
\bra \phi | \psi \ket_{L\!/\!R} = ( \phi \,|\, \Sig_{L\!/\!R}\, \psi) \:.
\eeq
We refer to~$\Sig_{L\!/\!R}$ as the {\bf{chiral signature operators}}. 
Taking the complex conjugate of the equation in~\eqref{SLRdef2} and using
that~$\chi_L^*=\chi_R$, we find that~\eqref{Sigprop} holds, where
the star denotes the adjoint in~$\Lin(\H_m)$.

We now define the chiral index as the Noether index of~$\Sig_L$
(sometimes called Fredholm index; for basics see for example~\cite[\S27.1]{lax}).
\begin{Def} \label{defind}
The fermionic signature operator is said to have finite chiral index if the operators of~$\Sig_L$
and~$\Sig_R$ both have a finite-dimensional kernel.
The {\bf{chiral index}} of the fermionic signature operator is defined by
\beq \label{ind}
\ind \Sig = \dim \ker \Sig_L - \dim \ker \Sig_R \:.
\eeq
\end{Def}

\section{Generalization to the Setting of Causal Fermion Systems} \label{seccfs}
Our starting point is a causal fermion system as introduced in~\cite{rrev}.
\begin{Def} {\em{
Given a complex Hilbert space~$(\H, \la .|. \ra_\H)$ (the {\em{``particle space''}})
and a parameter~$n \in \N$ (the {\em{``spin dimension''}}), we let~$\F \subset \Lin(\H)$ be the set of all
self-adjoint operators on~$\H$ of finite rank, which (counting with multiplicities) have
at most~$n$ positive and at most~$n$ negative eigenvalues. On~$\F$ we are given
a positive measure~$\rho$ (defined on a $\sigma$-algebra of subsets of~$\F$), the so-called
{\em{universal measure}}. We refer to~$(\H, \F, \rho)$ as a {\em{causal fermion system}}.
}}
\end{Def} \noindent
Starting from a Lorentzian spin manifold, one can construct a corresponding causal fermion system by
choosing~$\H$ as a suitable subspace of the solution space of the Dirac equation, 
forming the local correlation operators (possibly introducing an ultraviolet regularization) and defining~$\rho$
as the push-forward of the volume measure on~$\scrM$
(see~\cite[Section~4]{finite} or the examples in~\cite{topology}).
The advantage of working with a causal fermion system
is that the underlying space-time does not need to be a Lorentzian manifold, but it can be a more
general ``quantum space-time'' (for more details see~\cite{lqg}).

We now recall a few basic notions from~\cite{rrev}. 
On~$\F$ we consider the topology induced by the
operator norm~$\|A\| := \sup \{ \|A u \|_\H \text{ with } \| u \|_\H = 1 \}$.
For every~$x \in \F$
we define the {\em{spin space}}~$S_x$ by~$S_x = x(\H)$; it is a subspace of~$\H$ of dimension
at most~$2n$. On~$S_x$ we introduce the {\em{spin scalar product}} $\Sl .|. \Sr_x$ by
\beq \label{ssp}
\Sl u | v \Sr_x = -\la u | x u \ra_\H \qquad \text{(for all $u,v \in S_x$)}\:;
\eeq
it is an indefinite inner product of signature~$(p,q)$ with~$p,q \leq n$.
Moreover, we define {\em{space-time}}~$M$ as the support of the universal measure, $M = \text{supp}\, \rho$.
It is a closed subset of~$\F$.

In order to extend the chiral grading to causal fermion systems, we
assume for every~$x \in M$ an operator~$\pseudo(x) \in \Lin(\H)$ with the properties
\beq \label{pseudodef}
\pseudo(x)|_{S_x} \::\: S_x \rightarrow S_x \qquad \text{and} \qquad
x\, \pseudo(x) = -\pseudo(x)^*\, x \:.
\eeq
We define the operators~$\chi_{L\!/\!R}(x) \in \Lin(\H)$ again by~\eqref{chiLRdef}.
In order to explain the equations~\eqref{pseudodef}, we first note
that the right side of~\eqref{pseudodef} obviously vanishes on the
orthogonal complement of~$S_x$. Using furthermore that, by definition of the spin space,
the operator~$x$ is invertible on~$S_x$, we infer that
\[ \pseudo(x)|_{S_x^\perp} = 0 \:. \]
Moreover, the computation
\begin{align*}
\Sl \psi \,|\, \pseudo(x)\, \phi \Sr_x &= -\la \psi \,|\, x \,\pseudo(x)\, \phi \ra_\H = 
-\la \pseudo(x)^* x\, \psi \,|\, \phi \ra_\H \\
&\!\!\overset{\eqref{pseudodef}}{=}
\la x \,\pseudo(x)\, \psi \,|\, \phi \ra_\H = -\Sl \pseudo(x)\, \psi \,|\,  \phi \Sr_x
\end{align*}
(with~$\psi, \phi \in S_x$) shows that~$\pseudo(x) \in \Lin(S_x)$ is antisymmetric with respect to the
spin scalar product. Thus the first equation in~\eqref{gammaprop} again holds.
This implies that the adjoint of~$\chi_L(x)$ with respect to~$\Sl .|. \Sr_x$ equals~$\chi_R(x)$.
However, we point out that our assumptions~\eqref{pseudodef} do not imply that~$\pseudo(x)$
is idempotent (in the sense that~$\pseudo(x)^2|_{S_x}=\1_{S_x}$). Hence the analog
of the second equation in~\eqref{gammaprop} does not need to hold on a causal fermion system.
This property could be imposed in addition, but will not be needed here.
The last two relations in~\eqref{gammaprop} do not have an obvious correspondence on causal fermion systems,
and they will also not be needed in what follows.

We now have all the structures needed for defining the fermionic signature operator
and its chiral index.
Namely, replacing the scalar product in~\eqref{Sdef} by the scalar product on the
particle space~$\la  .|. \ra_\H$, we now demand in analogy to~\eqref{stip} and~\eqref{Sdef}
that the relation
\[ \la u | \Sig v \ra_\H = \int_M \Sl u | v \Sr_x\: d\rho(x) \]
should hold for all~$u, v \in \H$. Using~\eqref{ssp}, we find that the fermionic signature operator
is given by the integral
\[ \Sig = -\int_M x \: d\rho(x) \:. \]
Similarly, the left-handed signature operator can be introduced by
\beq \label{sigLint}
\Sig_L = -\int_M x \,\chi_L\: d\rho(x) \:.
\eeq
In the setting on a globally hyperbolic manifold, 
we had to assume that the manifold was $m$-finite and $\Gamma$-finite (see Definitions~\ref{defmfinite}
and~\ref{defGfinite}).
Now we need to assume correspondingly that the integral~\eqref{sigLint} converges.
For the sake of larger generality we prefer to work with weak convergence.

\begin{Def} The topological fermion system is {\bf{$\Sig_L$-bounded}} if the integral in~\eqref{sigLint}
converges weakly to a bounded operator, i.e.\ if there is an operator~$\Sig_L \in \Lin(\H)$
such that for all~$u,v \in \H$,
\[ -\int_M \la u \,|\, x \,\chi_L v \ra_\H \: d\rho(x) = \la u | \Sig_L v \ra_\H \:. \]
\end{Def} \noindent
Introducing the right-handed signature operator by~$\Sig_R := \Sig_L^*$,
we can define the {\bf{chiral index}} again by~\eqref{ind}.

\section{The Chiral Index in the Massless Odd Case} \label{secodd}
We return to the setting of Section~\ref{secindex}
and consider the special case that the mass vanishes and that the Dirac operator is odd,
\beq \label{massless}
m=0 \qquad \text{and} \qquad \pseudo \,\Dir = - \Dir\, \pseudo \:.
\eeq
In this case, the solution space of the Dirac equation is obviously invariant under~$\pseudo$,
\[ \pseudo \::\: \H_0 \rightarrow \H_0 \:. \]
Taking the adjoint with respect to the scalar product~\eqref{print} and noting that~$\pseudo$
anti-commutes with~$\slashed{\nu}$, one sees that~$\pseudo$ is symmetric on~$\H_0$.
Hence~$\chi_L$ and~$\chi_R$ are orthogonal projection operators,
giving rise to the orthogonal sum decomposition
\beq \label{HLR}
\H_0 = \H_L \oplus \H_R \qquad \text{with} \qquad \H_{L\!/\!R} := \chi_{L\!/\!R}\, \H_0\:.
\eeq
Moreover, the computation
\begin{align*}
\bra \chi_L \psi | \chi_{c'} \phi \ket_c &=
\int_\scrM \Sl \chi_L \psi \,|\, \chi_c \,\chi_{c'} \phi \Sr_x\: d\mu_\scrM \\
&= \int_\scrM \Sl \psi \,|\, \chi_R \,\chi_c \,\chi_{c'} \phi \Sr_x\: d\mu_\scrM
= \delta_{Rc} \:\delta_{cc'} \:\bra \psi | \phi \ket_c
\end{align*}
with~$c, c' \in \{L, R\}$ (and similarly for~$L$ replaced by~$R$) shows that~$\Sig$
maps the right-handed component to the left-handed component and vice versa.
Moreover, in a block matrix notation corresponding to the decomposition~\eqref{HLR},
the operators~$\Sig_L$ and~$\Sig_R$ have the simple form
\[ \Sig_L = \begin{pmatrix} 0 & 0 \\ A & 0 \end{pmatrix} \qquad \text{and} \qquad
\Sig_R = \begin{pmatrix} 0 & A^* \\ 0 & 0 \end{pmatrix} \]
with a bounded operator~$A : \H_L \rightarrow \H_R$.
As a consequence, both~$\Sig_L$ and~$\Sig_R$ have an infinite-dimensional kernel,
so that the index cannot be defined by~\eqref{ind}. This problem can easily be cured by
restricting the operators to the respective subspace~$\H_L$ and~$\H_R$.

\begin{Def} \label{defind0}
In the massless odd case~\eqref{massless}, the fermionic signature operator is said to have finite
chiral index if the operators~$\Sig_L|_{\H_L}$ and~$\Sig_R|_{\H_R}$ both have a finite-dimensional kernel.
We define the index~$\ind_0 \Sig$ by
\[ \ind_0 \Sig = \dim\ker (\Sig_L)|_{\H_L} - \dim \ker (\Sig_R)|_{\H_R} \:. \]
\end{Def}

\section{Homotopy Invariance} \label{secstable}
We first recall Dieudonn{\'e}'s general theorem on the homotopy invariance of the Noether index
(see for example~\cite[Theorem~27.1.5'']{lax}).
\begin{Thm} \label{thmhomotopy}
Let~$T(t) : U \rightarrow V$, $0 \leq t \leq 1$, be a one-parameter family of bounded linear operators
between Banach spaces~$U$ and~$V$ which is continuous in the norm topology. If
for every~$t \in [0,1]$ the vector spaces
\beq \label{kerfinite}
\text{$\ker T$ and~$V/T(\H)$ are both finite-dimensional}\:,
\eeq
then
\[ \ind T(0) = \ind T(1) \:, \]
where~$\ind T := \dim \ker(T) - \dim V/T(\H)$.
\end{Thm}

In most applications of this theorem, one knows from general arguments that the index
of~$T$ remains finite under homotopies (for example, in the prominent example of the
Atiyah-Singer index, this follows from elliptic estimates on a compact manifold).
For our chiral index, however, there is no general reason why the chiral index of~$\Sig$ should
remain finite. Indeed, the fermionic signature operator is bounded and typically has many eigenvalues
near zero. It may well happen that for a certain value of~$t$, an infinite number of these
eigenvalues becomes zero (for an explicit example see Example~\ref{exinstable} below).

Another complication when applying Theorem~\ref{thmhomotopy} to the fermionic signature operator is that the
image of~$\Sig_L$ does not need to be a closed subspace of our Hilbert space.
To explain the difficulty, we first consider the chiral index of Definition~\ref{defind}.
Using that~$\ker \Sig_R = \ker \Sig_L^* = \Sig_L(\H_m)^\perp$,
the assumption that the fermionic signature operator has finite chiral index can be restated
that the vector spaces~$\ker \Sig_L$ and~$\Sig_L(\H_m)^\perp$ are finite-dimensional subspaces
of~$\H_m$. Since
\[ \dim \Sig_L(\H_m)^\perp = \dim \H_m / \big( \overline{\Sig_L(\H_m)} \big) \:, \]
this implies that the {\em{closure}} of the image of~$\Sig_L$ has finite co-dimension.
If the image of~$\Sig_L$ were closed in~$\H_m$, the finiteness of the chiral index
would imply that the conditions~\eqref{kerfinite} hold if we set~$T=\Sig_L$ and~$U=V=\H_m$.
However, the image of~$\Sig_L$ will in general {\em{not}} be a closed subspace of~$\H_m$,
and in this case it is possible that the condition~\eqref{kerfinite} is violated for~$T=\Sig_L$
and~$U=V=\H_m$, although~$\Sig$ has finite chiral index (according to Definition~\ref{defind}).
In the massless odd case, the analogous problem occurs if we choose~$T=\Sig_L$,
$U=\H_L$ and~$V=\H_R$ (see Definition~\ref{defind0}).

Our method for making Theorem~\ref{thmhomotopy} applicable is to endow a subspace
of the Hilbert space with a finer topology, such that the image of~$\Sig_L$
lies in this subspace and is closed in this topology.

\begin{Thm} \label{thmstable}
Let~$\Sig(t) : \H_m \rightarrow \H_m$, $t \in [0,1]$, be a family of fermionic signature operators
with finite chiral index.
Let~$E$ be a Banach space together with an embedding~$\iota : E \hookrightarrow \H_m$
with the following properties:
\begin{itemize}
\item[(i)] For every~$t \in [0,1]$, the image of~$\Sig_L(t)$ lies in~$\iota(E)$, giving rise to the mapping
\beq \label{SigLE}
\Sig_L(t) \,:\, \H_m \rightarrow E \:.
\eeq
\item[(ii)] For every~$t \in [0,1]$, the image of the operator~$\Sig_L$, \eqref{SigLE}, is a closed subspace of~$E$.
\item[(iii)] The family~$\Sig_L(t) : \H_m \rightarrow E$ is continuous in the norm topology.
\end{itemize}
Then the chiral index is a homotopy invariant,
\[ \ind \Sig(0) = \ind \Sig(1)\:. \]
\end{Thm}

In the chiral odd case, the analogous result is stated as follows.
\begin{Thm} \label{thmstablem0}
Let~$\Sig(t) : \H_0 \rightarrow \H_0$, $t \in [0,1]$, be
a family of fermionic signature operators of finite chiral index in the massless odd case (see~\eqref{massless}).
Moreover, let~$E$ be a Banach space together with an embedding~$\iota : E \hookrightarrow \H_R$
such that the operator~$\Sig_L|_{\H_L} : \H_L \rightarrow \H_R$ has the following properties:
\begin{itemize}
\item[(i)] For every~$t \in [0,1]$, the image of~$\Sig_L(t)$ lies in~$\iota(E)$, giving rise to the mapping
\beq \label{SigLE2}
\Sig_L(t) \,:\, \H_L \rightarrow E \:.
\eeq
\item[(ii)] For every~$t \in [0,1]$, the image of the operator~$\Sig_L$, \eqref{SigLE2}, is a closed subspace of~$E$.
\item[(iii)] The family~$\Sig_L(t) : \H_L \rightarrow E$ is continuous in the norm topology.
\end{itemize}
Then the chiral index in the massless odd case is a homotopy invariant,
\[ \ind_0 \Sig(0) = \ind_0 \Sig(1)\:. \]
\end{Thm}
In Example~\ref{exstable} below, it will be explained how these theorems can be applied.

\section{Example: Shift Operators in the Setting of Causal Fermion Systems}
In the remainder of this paper we illustrate the previous constructions in several examples.
The simplest examples for fermionic signature operators with a non-trivial chiral index
can be given in the setting of causal fermion systems.
We let~$\H=\ell^2(\N)$ be the square-summable sequences with the scalar product
\[ \la u | v \ra_\H = \sum_{l=1}^\infty \overline{u_l} v_l \:. \]
For any~$k \in \N$ we define the operators~$x_k$ by
\[ (x_k \,u)_k = -u_{k+1} \:,\qquad (x_k \,u)_{k+1} = -u_k \:, \]
and all other components of~$x_k u$ vanish. Thus, writing the series in components,
\beq \label{xkdef}
x_k \,u = (\underbrace{0,\ldots, 0}_{\text{$k-1$ entries}}, -u_{k+1}, -u_k, 0, \ldots ) \:.
\eeq
Every operator~$x _k$ obviously has rank two with the non-trivial eigenvalues~$\pm 1$.
We let~$\mu$ be the counting measure on~$\N$ and~$\rho = x_*(\mu)$ the
push-forward measure of the mapping~$x : k \mapsto x_k \in \F \subset \Lin(\H)$.
We thus obtain a causal fermion system~$(\H, \F, \rho)$ of spin dimension one.

Next, we introduce the pseudoscalar operators~$\pseudo(x_k)$ by
\beq \label{Gkdef}
\pseudo(x_k) \,u = (\underbrace{0,\ldots, 0}_{\text{$k-1$ entries}}, u_k, -u_{k+1}, 0, \ldots ) \:.
\eeq
Obviously, these operators have the properties~\eqref{pseudodef}. Moreover,
\begin{eqnarray*}
x\, \chi_L(x_k)\, u = (0,\ldots, 0, & \!\!\!\!\!-u_{k+1}, \;\;0, & \!\!\!0, \ldots ) \\
x\, \chi_R(x_k)\, u = (0,\ldots, 0, & \:\;\;\;0, \;\;-u_k, & \!\!\!0, \ldots ) \:.
\end{eqnarray*}
Consequently, the operators
\beq \label{shiftsum}
\Sig_{L\!/\!R} = -\sum_{k=1}^\infty x\, \chi_L(x_k)
\eeq
take the form
\[ \Sig_L \,u = (u_2, u_3, u_4, \ldots) \:,\qquad \Sig_R \,u = (0, u_1, u_2, \ldots) \]
(note that the series in~\eqref{shiftsum} converges weakly; in fact it even converges strongly
in the sense that the series $\sum_k (x_k \chi_L u)$ converges in~$\H$ for every~$u \in \H$).
These are the usual shift operators, implying that
\[ \ind \Sig = 1 \:. \]

We finally remark that a general index~$p \in \N$ can be arranged by modifying~\eqref{xkdef}
and~\eqref{Gkdef} to
\begin{align*}
x_k \,u &= (\underbrace{0,\ldots, 0}_{\text{$k-1$ entries}}, -u_{k+p}, \!\!
\underbrace{0, \ldots, 0}_{\text{$p-1$ entries}}, \;-u_k, \;\:0, \ldots ) \\
\pseudo(x_k) \,u &= (\;\;\overbrace{0,\ldots, 0}\;\:, \;\;\;\;u_k,\;\;\; \overbrace{0, \ldots, 0}, \:-u_{k+p}, 0, \ldots ) \:.
\end{align*}
Moreover, a negative index can be arranged by exchanging the left- and right-handed components.

\section{Example: A Dirac Operator with~$\ind_0 \Sig \neq 0$} \label{secex1}
We now construct a two-dimensional space-time~$(\scrM,g)$ together with an odd Dirac operator~$\D$
such that the resulting fermionic signature operator in the massless case has a non-trivial chiral
index~$\ind_0$ (see Definition~\ref{defind0}).
We choose~$\scrM=(0, 2 \pi) \times S^1$ with coordinates~$t \in (0, 2 \pi)$ and~$\varphi \in [0, 2 \pi)$.
We begin with the flat Lorentzian metric
\beq \label{2dl}
ds^2 = dt^2 - d\varphi^2 \:.
\eeq
We consider two-component complex spinors, with the spin scalar product
\beq \label{ssprod}
\Sl \psi | \phi \Sr = \la \psi | \begin{pmatrix} 0 & 1 \\ 1 & 0 \end{pmatrix} \phi \ra_{\C^2}\:.
\eeq
We choose the pseudoscalar matrix as
\beq \label{pseudoc}
\pseudo = \begin{pmatrix} -1 & 0 \\ 0 & 1 \end{pmatrix} \:,
\eeq
so that
\beq \label{defchir}
\chi_L = \begin{pmatrix} 1 & 0 \\ 0 & 0 \end{pmatrix} \:,\qquad
\chi_R = \begin{pmatrix} 0 & 0 \\ 0 & 1 \end{pmatrix} \:.
\eeq
The space-time inner product~\eqref{stip} becomes
\beq \label{stiptorus}
\bra \psi|\phi \ket = \int_0^{2 \pi} \int_0^{2 \pi} \Sl \psi(t, \varphi) \,|\, \phi(t, \varphi) \Sr\:d\varphi\, dt \:.
\eeq

The Dirac operator~$\Dir$ should be chosen to be odd (see the right equation in~\eqref{massless}).
This means that~$\Dir$ has the matrix representation
\beq \label{Direx}
\Dir = \begin{pmatrix} 0 & \Dir_R \\ \Dir_L & 0 \end{pmatrix}
\eeq
with suitable operators~$\Dir_L$ and~$\Dir_R$.
In order for current conservation to hold, the Dirac operator should be symmetric 
with respect to the inner product~\eqref{stiptorus}. This implies that the operators~$\Dir_L$
and~$\Dir_R$ must both be symmetric,
\beq \label{DLRsymm}
\Dir_L^* = \Dir_L \:,\qquad \Dir_R^* = \Dir_R \:,
\eeq
where the star denotes the formal adjoint with respect to the scalar product on
the Hilbert space~$L^2(\scrM, \C)$. We consider the massless Dirac equation
\beq \label{Dir0}
\Dir \psi = 0 \:.
\eeq
The scalar product~\eqref{print} on the solutions takes the form
\beq \label{printtorus}
(\psi | \phi) = 2 \pi \int_0^{2 \pi} \la \psi(t,\varphi) | \phi(t, \varphi) \ra_{\C^2}\: d\varphi \:,
\eeq
giving rise to the Hilbert space~$(\H_0, (.|.))$. As a consequence of current conservation,
this scalar product is independent of the choice of~$t$.

We assume that the system is invariant under time translations and is a first order differential
operator in time. More precisely, we assume that
\beq \label{DirHam}
\Dir_{L\!/\!R} = i \partial_t - H_{L\!/\!R}
\eeq
with purely spatial operators~$H_{L\!/\!R}$, referred to as the left- and right-handed Hamiltonians.
Moreover, we assume that these Hamiltonians are homogeneous.
This implies that they can be diagonalized by plane waves,
\[ \Dir_c \:e^{i k \varphi} = \omega_{k,c}\: e^{i k \varphi} \qquad \text{with~$k \in \Z$ and~$c \in \{L, R\}$}\:. \]
As a consequence, the Dirac equation~\eqref{Dir0} can be solved by the plane waves
\beq \label{esols}
\e_{k, L} = \frac{1}{2 \pi}\: e^{-i \omega_{k, L} t + i k \varphi} \begin{pmatrix} 1 \\ 0 \end{pmatrix} \:,\qquad
\e_{k, R} = \frac{1}{2 \pi}\: e^{-i \omega_{k, R} t + i k \varphi} \begin{pmatrix} 0 \\ 1 \end{pmatrix} \:.
\eeq
The vectors~$(\e_{k, c})_{k \in \Z, c \in \{L, R\}}$ form an orthonormal basis of the
Hilbert space~$\H_0$.
We remark that the Dirac operator of the Minkowski vacuum is obtained by choosing
\[ H_L = i \partial_\varphi \:,\qquad H_R = -i \partial_\varphi \]
(see for example~\cite{drum} or~\cite[Section~7.2]{topology}).
In this case, $\omega_{k, L\!/\!R} = \mp k$. More generally, choosing~$\Dir_c$ as a
homogeneous differential operator of first order, the eigenvalues~$\omega_{k,c}$ are linear in~$k$.
Here we do not want to assume that the operators~$\Dir_c$ are differential operators.
Then the eigenvalues~$\omega_{k, L}$ and~$\omega_{k,R}$
can be chosen arbitrarily and independently, except for the constraint coming from the
symmetry~\eqref{DLRsymm} that these eigenvalues must be real.

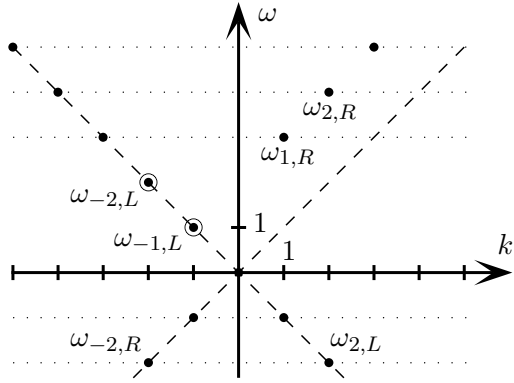
\begin{figure}
\scalebox{1} 
{
\begin{pspicture}(0,-2.52)(6.92,2.52)
\psdots[dotsize=0.24,fillstyle=solid,dotstyle=o](1.86,0.1)
\psdots[dotsize=0.24,fillstyle=solid,dotstyle=o](2.46,-0.5)
\usefont{T1}{ptm}{m}{n}
\rput(6.61,-0.715){$k$}
\psline[linewidth=0.04cm,arrowsize=0.3cm 1.0,arrowlength=1.5,arrowinset=0.5]{->}(0.04,-1.1)(6.7,-1.1)
\psline[linewidth=0.04cm,arrowsize=0.3cm 1.0,arrowlength=1.5,arrowinset=0.5]{->}(3.06,-2.5)(3.06,2.5)
\usefont{T1}{ptm}{m}{n}
\rput(3.45,2.325){$\omega$}
\psline[linewidth=0.02cm,linestyle=dashed,dash=0.16cm 0.16cm](1.66,-2.5)(6.06,1.9)
\psline[linewidth=0.02cm,linestyle=dashed,dash=0.16cm 0.16cm](0.06,1.9)(4.46,-2.5)
\psdots[dotsize=0.12](3.66,0.7)
\psdots[dotsize=0.12](3.06,-1.1)
\psdots[dotsize=0.12](2.46,-0.5)
\psdots[dotsize=0.12](1.26,0.7)
\psdots[dotsize=0.12](0.06,1.9)
\psdots[dotsize=0.12](4.26,-2.3)
\psdots[dotsize=0.12](2.46,-1.7)
\psdots[dotsize=0.12](1.86,-2.3)
\psdots[dotsize=0.12](1.86,0.1)
\psdots[dotsize=0.12](0.66,1.3)
\psdots[dotsize=0.12](4.26,1.3)
\psdots[dotsize=0.12](4.86,1.9)
\psdots[dotsize=0.12](3.66,-1.7)
\psline[linewidth=0.04cm](3.66,-1.0)(3.66,-1.2)
\psline[linewidth=0.04cm](4.26,-1.0)(4.26,-1.2)
\psline[linewidth=0.04cm](4.86,-1.0)(4.86,-1.2)
\psline[linewidth=0.04cm](2.46,-1.0)(2.46,-1.2)
\psline[linewidth=0.04cm](1.86,-1.0)(1.86,-1.2)
\psline[linewidth=0.04cm](1.26,-1.0)(1.26,-1.2)
\psline[linewidth=0.04cm](0.66,-1.0)(0.66,-1.2)
\psline[linewidth=0.04cm](6.06,-1.0)(6.06,-1.2)
\psline[linewidth=0.04cm](5.46,-1.0)(5.46,-1.2)
\psline[linewidth=0.04cm](0.06,-1.0)(0.06,-1.2)
\psline[linewidth=0.04cm](3.16,-0.5)(2.96,-0.5)
\psline[linewidth=0.02cm,linestyle=dotted,dotsep=0.16cm](0.06,0.7)(6.26,0.7)
\psline[linewidth=0.02cm,linestyle=dotted,dotsep=0.16cm](0.06,1.3)(6.26,1.3)
\psline[linewidth=0.02cm,linestyle=dotted,dotsep=0.16cm](0.06,1.9)(6.26,1.9)
\psline[linewidth=0.02cm,linestyle=dotted,dotsep=0.16cm](0.06,-1.7)(6.26,-1.7)
\psline[linewidth=0.02cm,linestyle=dotted,dotsep=0.16cm](0.06,-2.3)(6.26,-2.3)
\usefont{T1}{ptm}{m}{n}
\rput(3.35,-0.435){$1$}
\usefont{T1}{ptm}{m}{n}
\rput(1.87,-0.715){$\omega_{-1,L}$}
\usefont{T1}{ptm}{m}{n}
\rput(1.29,-0.135){$\omega_{-2,L}$}
\usefont{T1}{ptm}{m}{n}
\rput(3.73,-0.815){$1$}
\usefont{T1}{ptm}{m}{n}
\rput(3.72,0.425){$\omega_{1,R}$}
\usefont{T1}{ptm}{m}{n}
\rput(4.63,-2.055){$\omega_{2,L}$}
\usefont{T1}{ptm}{m}{n}
\rput(4.28,1.025){$\omega_{2,R}$}
\usefont{T1}{ptm}{m}{n}
\rput(1.3,-2.055){$\omega_{-2,R}$}
\end{pspicture} 
}
\caption{The eigenvalues~$\omega_{k, L\!/\!R}$ in the case~$p=2$.}
\label{figdisperse}
\end{figure}

More specifically, for a given parameter~$p \in \N$ we choose
\beq \label{omegaex}
\omega_{k,L} = -k \qquad \text{and} \qquad
\omega_{k, R} = \left\{ \begin{array}{cl} k & \text{if~$k \leq 0$} \\
k+p & \text{if~$k > 0$}
\end{array} \right.
\eeq
(see Figure~\ref{figdisperse}).
Then the space-time inner product of the basis vectors~$(\e_{k, c})_{k \in \Z, c \in \{L, R\}}$
is computed by
\begin{align*}
\bra \e_{k,L} | \e_{k', L} \ket &= 0 = \bra \e_{k,R} | \e_{k', R} \ket \\
\bra \e_{k,R} | \e_{k', L} \ket 
&= \frac{1}{2 \pi} \: \delta_{k,k'} \int_0^{2 \pi} e^{i (\omega_{k,R} - \omega_{k,L}) t} \,dt =
\delta_{k,k'} \: \delta_{\omega_{k,R}, \;\omega_{k,L}} = \delta_{k,0} \: \delta_{k',0} \:.
\end{align*}
We conclude~$\Sig$ does not have finite chiral index.

In order to obtain a non-trivial index, we need to modify our example.
The idea is to change the space-time inner product in such a way that
the inner product between two different plane-wave solutions with
the same frequencies becomes non-zero. As a consequence,
the corresponding pair of plane-wave solutions will disappear from the kernel.
The only vectors which remain in the kernel are those which do not have a partner
for pairing, so that
\[ \ker \Sig_L|_{\H_L} = \Span \big( \e_{-1,L}, \ldots, \e_{-p,L} \big) \:,\qquad
\ker \Sig_R|_{\H_R} = \{0\}\:. \]
(see again Figure~\ref{figdisperse}, where the pairs are indicated by horizontal dashed lines,
whereas the vectors in the kernel correspond to the circled dots).
Generally speaking, the method to modify the space-time inner product for states with
the same frequency is to insert a potential into the Dirac equation which
is time-independent but has a non-trivial spatial dependence.
It is most convenient to work with a {\em{conformal transformation}}.
Thus we go over from the Minkowski metric~\eqref{2dl} to the conformally flat metric
\beq \label{conform}
d\tilde{s}^2 = f(\varphi)^2 \left( dt^2 - d\varphi^2 \right) \:,
\eeq
where~$f \in C^\infty(\R/(2 \pi \Z))$ is a non-negative, smooth, $2 \pi$-periodic function.
The conformal invariance of the Dirac equation (for details see for example~\cite[Section~8.1]{topology}
and the references therein) states in our situation that the Dirac operator transforms as
\beq
\tilde{\Dir} = f^{-\frac{3}{2}} \,\Dir\, f^{\frac{1}{2}} \:, \label{Dirconf}
\eeq
so that
\[ \tilde{\Dir} = \begin{pmatrix} 0 & \tilde{\Dir}_R \\ \tilde{\Dir}_L & 0 \end{pmatrix}
\qquad \text{with} \qquad
\tilde{\Dir}_{L\!/\!R} = f^{-\frac{3}{2}} \,\Dir_{L\!/\!R}\, f^{\frac{1}{2}} \:. \]
The solutions of the massless Dirac equation are modified simply by a conformal factor,
\beq \label{psiconf}
\tilde{\psi} =  f^{-\frac{1}{2}}\: \psi \:.
\eeq
The space-time inner product~\eqref{stiptorus} and the scalar product~\eqref{printtorus} transform to
\begin{align}
\bra \tilde{\psi} | \tilde{\phi} \ket
&= \int_0^{2 \pi} \int_0^{2 \pi} \Sl \tilde{\psi}(t, \varphi) \,|\, \tilde{\phi}(t, \varphi) \Sr\:
f(\varphi)^2 \,d\varphi\, dt \nonumber \\
&= \int_0^{2 \pi} \int_0^{2 \pi} \Sl \psi(t, \varphi) \,|\, \phi(t, \varphi) \Sr\:
f(\varphi) \,d\varphi\, dt \label{stiptrans} \\
(\tilde{\psi} | \tilde{\phi}) &= 2 \pi \int_0^{2 \pi} \la \tilde{\psi}(t,\varphi) | \tilde{\phi}(t, \varphi) \ra_{\C^2}\: 
f(\varphi) \:d\varphi \nonumber \\
&= 2 \pi \int_0^{2 \pi} \la \psi(t,\varphi) | \phi(t, \varphi) \ra_{\C^2}\: 
d\varphi = (\psi | \phi) \:. \label{printtrans}
\end{align}
To understand these transformation laws, one should keep in mind that the spin scalar product
remains unchanged under conformal transformations. The same is true for the 
integrand~$\Sl \psi | \nuslsh \phi \Sr_x$ of the scalar product~\eqref{print}, because
the operator~$\slashed{\nu}$ is normalized by~$\slashed{\nu}^2=\1$.

From~\eqref{printtrans} we conclude that the scalar product does not change under conformal
transformations. In particular, the conformally transformed plane-wave solutions
\beq \label{vertical}
\tilde{\e}_{k, L\!/\!R} = f(\varphi)^{-\frac{1}{2}}\: \e_{k, L\!/\!R}
\eeq
are an orthonormal basis of~$\tilde{\H}_0$. The space-time inner product~\eqref{stiptrans}, however,
involves a conformal factor~$f(\varphi)$. As a consequence, the space-time inner product
of the basis vectors~$(\tilde{\e}_{k, c})_{k \in \Z, c \in \{L, R\}}$ can be computed by
\begin{align*}
\bra \tilde{\e}_{k,L} | \tilde{\e}_{k', L} \ket &= 0 = \bra \tilde{\e}_{k,R} | \tilde{\e}_{k', R} \ket \\
\bra \tilde{\e}_{k,R} | \tilde{\e}_{k', L} \ket &= \int_0^{2 \pi} dt \int_0^{2 \pi} f(\varphi)\, d\varphi\; 
\Sl \tilde{\e}_{k,R}(t,\varphi) \,|\, \tilde{\e}_{k', L}(t,\varphi) \Sr \\
&= \frac{1}{2 \pi} \:\delta_{\omega_{k,R}, \,\omega_{k',L}}
\int_0^{2 \pi} f(\varphi)\: e^{-i (k-k') \varphi}\, d\varphi
= \frac{1}{2 \pi} \:\delta_{\omega_{k,R}, \omega_{k',L}}\: \hat{f}_{k-k'}\:,
\end{align*}
where~$\hat{f}_k$ is the $k^\text{th}$ Fourier coefficient of~$f$,
\[ f(\varphi) = \frac{1}{2 \pi} \sum_{k \in \Z} \hat{f}_k\: e^{i k \varphi} \:. \]

Using the explicit form of the frequencies~\eqref{omegaex}, we obtain the following
invariant subspaces and corresponding matrix representations of~$\Sig$,
\begin{align*}
\hat{\Sig}|_{\Span(\tilde{\e}_{-k, L}, \tilde{\e}_{k, R})} &=
\frac{1}{2 \pi} \begin{pmatrix} 0 & \overline{\hat{f}_{2k}} \\ \hat{f}_{2k}  & 0 \end{pmatrix} 
&&\hspace*{-0.8cm} \text{if~$k \leq 0$} \\
\hat{\Sig}|_{\Span(\tilde{\e}_{-k-p, L}, \tilde{\e}_{k, R})} &=
\frac{1}{2 \pi} \begin{pmatrix} 0 & \overline{\hat{f}_{2k+p}} \\ \hat{f}_{2k+p}  & 0 \end{pmatrix} 
&&\hspace*{-0.8cm} \text{if~$k>p$} \\
\hat{\Sig}|_{\Span(\tilde{\e}_{-1, L}, \tilde{\e}_{-p, L})} &= 0 \:.
\end{align*}
In particular, we can read off the chiral index:
\begin{Prp} \label{prpindex} Assume that almost all Fourier coefficients~$\hat{f}_k$ 
of the conformal function in~\eqref{conform} are non-zero.
Then the fermionic signature operator in the massless odd case has finite chiral
index (see Definition~\ref{defind0}) and~$\ind_0 \Sig = p$.
\end{Prp}

We finally compute the Dirac operator in position space. The dispersion relations in~\eqref{omegaex}
are realized by the operators
\begin{align*}
\Dir_L &= i (\partial_t - \partial_\varphi) \\
\Dir_R &= i (\partial_t + \partial_\varphi) + \B \:,
\end{align*}
where~$\B$ is the spatial integral operator
\[ \big( \B \psi \big)(t,\varphi) = \int_0^{2 \pi} \B(\varphi, \varphi')\: \psi(t, \varphi')\: d\varphi' \]
with the distributional integral kernel
\[ \B(\varphi, \varphi') = -\frac{p}{2 \pi} \sum_{k=1}^\infty e^{i k (\varphi-\varphi')} \nonumber \\
= -\frac{p}{2}\: \delta(\varphi-\varphi') -\frac{p}{2 \pi}\:\frac{\text{PP}}{e^{-i(\varphi-\varphi')}-1} \:. \]
Hence, choosing the Dirac matrices as
\beq \label{gammadef}
\gamma^0 = \begin{pmatrix} 0 & 1 \\ 1 & 0 \end{pmatrix} \:,\qquad
\gamma^1 = \begin{pmatrix} 0 & 1 \\ -1 & 0 \end{pmatrix}
\eeq
and using~\eqref{defchir}, we obtain
\beq \label{Diracform}
\Dir = i \gamma^0 \partial_t + i \gamma^1 \partial_\varphi
+ \gamma^1 \chi_R \, \B \:.
\eeq
Performing the conformal transformation~\eqref{Dirconf}, we finally obtain
\begin{align}
\big(\tilde{\Dir} \psi)(t, \varphi) &= \frac{i}{f(\varphi)} \left( \gamma^0 \partial_t + \gamma^1 \partial_\varphi 
+ \frac{f'(\varphi)}{2 f(\varphi)} -\frac{p}{2} \: \gamma^1 \chi_R \right) \psi(t, \varphi) \label{Dir1} \\
&\qquad -\frac{p}{2 \pi}\:  \frac{\gamma^1 \chi_R}{f(\varphi)^\frac{3}{2}}
\int_0^{2 \pi} \frac{\text{PP}}{e^{-i(\varphi-\varphi')}-1}\:
\psi(t, \varphi')\: \sqrt{f(\varphi')}\: d\varphi' \:. \label{Dir2}
\end{align}
Thus~\eqref{Dir1} is the Dirac operator in the Lorentzian metric~\eqref{conform}
with a constant right-handed potential. Moreover, the summand~\eqref{Dir2} is a nonlocal integral
operator involving a singular integral kernel.

This example shows that the index of Proposition~\ref{prpindex} in general does not encode
the topology of space-time, because for a fixed space-time topology the index can take
any integer value. The way we understand the index is that it gives topological information
on the singular behavior of the potential in the Dirac operator.

\section{Example: A Dirac Operator with~$\ind \Sig \neq 0$} \label{secex2}
We now construct an example of a fermionic signature operator for which the
index~$\ind \Sig$ of Definition~\ref{defind} is non-trivial.
To this end, we want to modify the example of the previous section.
The major difference to the previous setting is that the Hilbert space~$\H_m$
does not have a decomposition into two subspaces~$\H_L$ and~$\H_R$, making
it necessary to consider the operators~$\Sig_L$ and~$\Sig_R$ as operators on the whole
solution space~$\H_m$. Our first task is to remove the infinite-dimensional kernels of the operators~$\Sig_L$
and~$\Sig_R$. This can typically be achieved by perturbing the Dirac operator, for example by introducing
a rest mass. The second and more substantial modification is to arrange that the
operators~$\Sig_L$ and~$\Sig_R$ have {\em{infinite-dimensional invariant subspaces}}.
This is needed for the following reason: In the example of the previous section,
the operator~$\Sig_L|_{\H_L} : \H_L \rightarrow \H_R$ mapped one Hilbert
space to another Hilbert space. Therefore, we obtained a non-trivial index simply by arranging
that the operator~$\Sig_L|_{\H_L}$ gives a non-trivial ``pairing'' of vectors of~$\H_L$ with
vectors of~$\H_R$ (as indicated in Figure~\ref{figdisperse} by the horizontal dashed lines).
In particular, if considered as an operator on~$\H_0$, the operator~$\Sig_L$ had at most two-dimensional
invariant subspaces.
For the chiral index of Definition~\ref{defind}, however, we have only one Hilbert space~$\H_m$
to our disposal, so that the operator~$\Sig_L : \H_m \rightarrow \H_m$ is an endomorphism of~$\H_m$.
As a consequence, the chiral index is trivial whenever~$\H_m$ splits into a direct sum
of finite-dimensional subspaces which are invariant under~$\Sig_L$
(because on each invariant subspace, the index is trivial due to the
rank-nullity theorem of linear algebra).

\begin{figure}
\scalebox{1} 
{
\begin{pspicture}(0,-2.175)(7.06,2.18)
\psdots[dotsize=0.24,fillstyle=solid,dotstyle=o](3.22,0.06)
\usefont{T1}{ptm}{m}{n}
\rput(6.23,-0.175){$k$}
\psline[linewidth=0.04cm,arrowsize=0.3cm 1.0,arrowlength=1.5,arrowinset=0.5]{->}(1.62,-0.54)(6.42,-0.54)
\psline[linewidth=0.04cm,arrowsize=0.3cm 1.0,arrowlength=1.5,arrowinset=0.5]{->}(3.82,-2.14)(3.82,2.16)
\usefont{T1}{ptm}{m}{n}
\rput(4.21,1.925){$\omega$}
\psline[linewidth=0.02cm,linestyle=dashed,dash=0.16cm 0.16cm](2.42,-1.94)(6.02,1.66)
\psline[linewidth=0.02cm,linestyle=dashed,dash=0.16cm 0.16cm](1.62,1.66)(5.22,-1.94)
\psdots[dotsize=0.12](4.42,0.66)
\psdots[dotsize=0.12](3.82,-0.54)
\psdots[dotsize=0.12](3.22,0.06)
\psdots[dotsize=0.12](2.02,1.26)
\psdots[dotsize=0.12](5.02,-1.74)
\psdots[dotsize=0.12](3.22,-1.14)
\psdots[dotsize=0.12](2.62,-1.74)
\psdots[dotsize=0.12](2.62,0.66)
\psdots[dotsize=0.12](5.02,1.26)
\psdots[dotsize=0.12](4.42,-1.14)
\psline[linewidth=0.04cm](4.42,-0.44)(4.42,-0.64)
\psline[linewidth=0.04cm](5.02,-0.44)(5.02,-0.64)
\psline[linewidth=0.04cm](5.62,-0.44)(5.62,-0.64)
\psline[linewidth=0.04cm](3.22,-0.44)(3.22,-0.64)
\psline[linewidth=0.04cm](2.62,-0.44)(2.62,-0.64)
\psline[linewidth=0.04cm](2.02,-0.44)(2.02,-0.64)
\psline[linewidth=0.04cm](3.92,0.06)(3.72,0.06)
\usefont{T1}{ptm}{m}{n}
\rput(4.11,0.125){$1$}
\usefont{T1}{ptm}{m}{n}
\rput(1.44,1.245){$\tilde{\e}_{-3,L}$}
\usefont{T1}{ptm}{m}{n}
\rput(4.49,-0.255){$1$}
\psline[linewidth=0.03cm,tbarsize=0.07055555cm 5.0]{|*-|*}(2.02,1.46)(2.02,1.06)
\psline[linewidth=0.03cm,tbarsize=0.07055555cm 5.0]{|*-|*}(5.02,1.46)(5.02,1.06)
\psline[linewidth=0.03cm,tbarsize=0.07055555cm 5.0]{|*-|*}(4.42,0.86)(4.42,0.46)
\psline[linewidth=0.03cm,tbarsize=0.07055555cm 5.0]{|*-|*}(2.62,0.86)(2.62,0.46)
\psline[linewidth=0.03cm,tbarsize=0.07055555cm 5.0]{|*-|*}(3.22,0.26)(3.22,-0.14)
\psline[linewidth=0.03cm,tbarsize=0.07055555cm 5.0]{|*-|*}(4.42,-0.94)(4.42,-1.34)
\psline[linewidth=0.03cm,tbarsize=0.07055555cm 5.0]{|*-|*}(3.22,-0.94)(3.22,-1.34)
\psline[linewidth=0.03cm,tbarsize=0.07055555cm 5.0]{|*-|*}(2.62,-1.54)(2.62,-1.94)
\psline[linewidth=0.03cm,tbarsize=0.07055555cm 5.0]{|*-|*}(5.02,-1.54)(5.02,-1.94)
\psbezier[linewidth=0.03,arrowsize=0.08cm 3.0,arrowlength=1.2,arrowinset=0.4]{->}(2.22,1.26)(3.02,1.54)(3.92,1.7)(4.82,1.26)
\psbezier[linewidth=0.03,arrowsize=0.08cm 3.0,arrowlength=1.2,arrowinset=0.4]{->}(5.22,1.16)(5.92,0.2)(5.9,-1.0)(5.2,-1.66)
\psbezier[linewidth=0.03,arrowsize=0.08cm 3.0,arrowlength=1.2,arrowinset=0.4]{->}(4.84,-1.78)(4.12,-2.1)(3.7,-2.16)(2.8,-1.78)
\psbezier[linewidth=0.03,arrowsize=0.08cm 3.0,arrowlength=1.2,arrowinset=0.4]{->}(2.44,-1.7)(1.74,-1.0)(1.72,-0.06)(2.48,0.64)
\psbezier[linewidth=0.03,arrowsize=0.08cm 3.0,arrowlength=1.2,arrowinset=0.4]{->}(2.8,0.68)(3.24,0.88)(3.66,1.0)(4.24,0.68)
\psbezier[linewidth=0.03,arrowsize=0.08cm 3.0,arrowlength=1.2,arrowinset=0.4]{->}(4.58,0.6)(5.04,-0.02)(4.98,-0.58)(4.58,-1.1)
\psbezier[linewidth=0.03,arrowsize=0.08cm 3.0,arrowlength=1.2,arrowinset=0.4]{->}(4.28,-1.18)(4.1,-1.3)(3.68,-1.36)(3.4,-1.18)
\psbezier[linewidth=0.03,arrowsize=0.08cm 3.0,arrowlength=1.2,arrowinset=0.4]{->}(3.06,-1.14)(2.72,-0.7)(2.8,-0.32)(3.04,0.04)
\usefont{T1}{ptm}{m}{n}
\rput(5.64,-1.875){$\tilde{\e}_{2,L}$}
\usefont{T1}{ptm}{m}{n}
\rput(5.43,1.645){$\tilde{\e}_{2,R}$}
\psline[linewidth=0.04cm,tbarsize=0.07055555cm 5.0]{|*-|*}(3.82,-0.34)(3.82,-0.74)
\end{pspicture} 
}
\caption{The action of~$\Sig_L$ on the transformed plane-wave solutions in the case~$p=1$.}
\label{figsnail}
\end{figure}
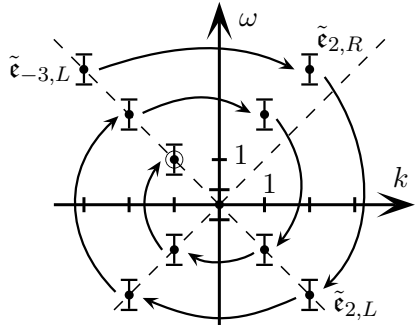

The following example is designed with the aim of showing in explicit detail that the
index is non-zero. Our starting point are the plane-wave solutions~\eqref{esols}
with the frequencies according to~\eqref{omegaex} with~$p \in \N$.
In Figure~\ref{figsnail} the transformed plane-wave solutions~$\tilde{\e}_{k,c}$ 
(where the transformation from~$\e_{k,c}$ to~$\tilde{\e}_{k,c}$ will be explained below) are
arranged according to their frequencies and momenta on a lattice.
We shall construct the operator~$\Sig_L$ in such a way that
these plane-wave solutions are mapped to each other as indicated by the arrows.
Thus similar to a shift operator, $\Sig_L$ maps the basis vectors to each other ``spiraling in,''
implying that the vector~$\tilde{\e}_{-1,L}$ (depicted with the circled dot) is in the kernel of~$\Sig_L$.
Likewise, the operator~$\Sig_R$ acts like a ``spiraling out'' shift vector, so that it is injective.
In this way, we arrange that~$\ind \Sig = 1$. Similarly, in the case~$p>1$ we shall obtain~$p$
spirals, so that~$\ind \Sig=p$.

Before entering the detailed construction, we point out that our method
is driven by the wish that the example should be explicit and that the kernels of the
chiral signature operators should be given in closed form.
This makes it necessary to introduce a Dirac operator which seems somewhat artificial.
In particular, instead of introducing a rest mass,
we arrange a mixing of the left- and right-handed components
using a time-dependent vectorial gauge transformation.
Moreover, we again work with a conformal transformation with a carefully adjusted
spatial and time dependence. We consider these special features merely as a
requirement needed in order to make the computations as simple as possible.
In view of the stability result of Theorem~\ref{thmstable}, we expect that the index
is also non-trivial in more realistic examples involving a rest mass and less fine-tuned potentials.
But probably, this goes at the expense of longer computations or less explicit arguments.

We begin on the cylinder~$\scrM=(0, 6 \pi) \times S^1$, again with the Minkowski metric~\eqref{2dl}
and two-component spinors endowed with the spin scalar product~\eqref{ssprod}.
The space-time inner product~\eqref{stip} becomes
\beq \label{stiptoru2}
\bra \psi|\phi \ket = \int_0^{6 \pi} \int_0^{2 \pi} \Sl \psi(t, \varphi) \,|\, \phi(t, \varphi) \Sr\: d\varphi \:dt \:,
\eeq
whereas the scalar product on solutions of the Dirac equation is again given by~\eqref{printtorus}.
We again consider the massless Dirac equation~\eqref{Dir0}
with the Dirac operator~\eqref{Direx} and the left- and right-handed
operators according to~\eqref{DirHam}. Moreover, we again assume that the operators~$\Dir_{L\!/\!R}$
have the plane-wave solutions~\eqref{esols} with frequencies~\eqref{omegaex}.
For a fixed real parameter~$\nu \neq 0$, we consider the transformation
\beq \label{Utdef}
U(t)  = \frac{\1 + i \nu \gamma^0 \cos t/3}{\sqrt{1 + \nu^2 \cos^2 t/3}} = 
\frac{1}{\sqrt{1 + \nu^2 \cos^2 t/3}}
\begin{pmatrix} 1 & i \nu \cos t/3 \\ i \nu \cos t/3 & 1 \end{pmatrix} \:.
\eeq
Obviously, $U(t) \in \U(2)$ is a unitary matrix. Moreover, it commutes with~$\gamma^0$,
implying that it is also unitary with respect to the spin scalar product.
As a consequence, the transformation~$U(t)$ is unitary both on the Hilbert space~$\H_0$
and with respect to the inner product~\eqref{stiptoru2}.
Next, we again consider a conformal transformation~\eqref{Dirconf} and~\eqref{psiconf},
but now with a conformal function~$f(t, \varphi)$ which depends on space and time.
Thus we set
\beq \label{Dirdef}
\tilde{\Dir} = f^{-\frac{3}{2}} U \,\Dir\, U^* f^{\frac{1}{2}} \qquad \text{and} \qquad
\tilde{\psi} = f^{-\frac{1}{2}}\: U \,\psi \:.
\eeq
Similar to~\eqref{stiptrans} and~\eqref{printtrans}, the inner products transform to
\[ \bra \tilde{\psi} | \tilde{\phi} \ket
= \int_0^{6 \pi} \int_0^{2 \pi} \Sl \psi(t, \varphi) \,|\, \phi(t, \varphi) \Sr\:
f(t, \varphi)\, d\varphi\,dt \qquad \text{and} \qquad
(\tilde{\psi} | \tilde{\phi}) = (\psi | \phi) \:. \]
In particular, the transformed plane wave solutions~$\tilde{e}_{k,c}$ are an orthonormal
basis of~$\H_0$.
Keeping in mind that the chiral projectors in~\eqref{stipLR} do {\em{not}} commute with~$U$,
we obtain
\[ \bra \tilde{\psi} |\tilde{\phi} \ket_L
= \int_0^{6 \pi} \int_0^{2 \pi} \Sl U(t)\, \psi(t, \varphi) \,|\, \chi_L\, U(t)\, \phi(t, \varphi) \Sr\:
f(t, \varphi)\, d\varphi\,dt \]
and thus, in view of~\eqref{SLRdef2},
\beq \label{SL1}
(\tilde{\e}_{k,c} \,|\, \Sig_L \,\tilde{\e}_{k',c'}) = 
\int_0^{6 \pi} \int_0^{2 \pi} \Sl U \e_{k,c} \,|\, \chi_L U \e_{k',c'} \Sr\:
f(t, \varphi)\, d\varphi\,dt \:.
\eeq
In order to get rid of the square roots in~\eqref{Utdef}, it is most convenient to set
\beq \label{lambdadef}
V(t) = \begin{pmatrix} 1 & i \nu \cos t/3 \\ i \nu \cos t/3 & 1 \end{pmatrix}
\qquad \text{and} \qquad \mu(t, \varphi) = \frac{f(t, \varphi)}{1 + \nu^2 \cos^2 t/3} \:.
\eeq
Then~\eqref{SL1} simplifies to
\beq \label{SL2}
(\tilde{\e}_{k,c} \,|\, \Sig_L \,\tilde{\e}_{k',c'}) = 
\int_0^{6 \pi} \int_0^{2 \pi} \Sl V \e_{k,c} \,|\, \chi_L V \e_{k',c'} \Sr\:
\mu(t, \varphi)\, d\varphi\,dt \:.
\eeq

\begin{figure}
\scalebox{1} 
{
\begin{pspicture}(0,-0.97)(10.51,0.97)
\psdots[dotsize=0.2](4.4,-0.15)
\psline[linewidth=0.04cm,tbarsize=0.07055555cm 5.0]{|*-|*}(5.6,0.45)(5.6,-0.75)
\psdots[dotsize=0.2](5.6,-0.15)
\usefont{T1}{ptm}{m}{n}
\rput(4.01,-0.125){$V$}
\usefont{T1}{ptm}{m}{n}
\rput(4.98,-0.145){$=$}
\usefont{T1}{ptm}{m}{n}
\rput(4.39,0.215){$L$}
\usefont{T1}{ptm}{m}{n}
\rput(6.19,-0.145){$L \;\;,$}
\usefont{T1}{ptm}{m}{n}
\rput(6.0,-0.745){$R$}
\usefont{T1}{ptm}{m}{n}
\rput(6.0,0.455){$R$}
\usefont{T1}{ptm}{m}{n}
\rput(2.01,-0.145){$\omega$}
\usefont{T1}{ptm}{m}{n}
\rput(1.68,0.455){$\omega+\frac{1}{3}$}
\usefont{T1}{ptm}{m}{n}
\rput(1.64,-0.745){$\omega-\frac{1}{3}$}
\psline[linewidth=0.04cm,arrowsize=0.04cm 4.0,arrowlength=2.0,arrowinset=0.4]{->}(2.4,-0.95)(2.4,0.95)
\psline[linewidth=0.04cm](2.3,-0.15)(2.5,-0.15)
\psline[linewidth=0.04cm](2.3,0.45)(2.5,0.45)
\psline[linewidth=0.04cm](2.3,-0.75)(2.5,-0.75)
\psline[linewidth=0.02cm,linestyle=dotted,dotsep=0.16cm](2.4,0.45)(5.5,0.45)
\psline[linewidth=0.02cm,linestyle=dotted,dotsep=0.16cm](2.4,-0.15)(3.6,-0.15)
\psline[linewidth=0.02cm,linestyle=dotted,dotsep=0.16cm](2.3,-0.75)(5.6,-0.75)
\usefont{T1}{ptm}{m}{n}
\rput(7.61,-0.125){$V$}
\psdots[dotsize=0.2](8.0,-0.15)
\usefont{T1}{ptm}{m}{n}
\rput(8.58,-0.145){$=$}
\psline[linewidth=0.04cm,tbarsize=0.07055555cm 5.0]{|*-|*}(9.2,0.45)(9.2,-0.75)
\psdots[dotsize=0.2](9.2,-0.15)
\usefont{T1}{ptm}{m}{n}
\rput(9.6,-0.145){$R$}
\usefont{T1}{ptm}{m}{n}
\rput(9.59,-0.745){$L$}
\usefont{T1}{ptm}{m}{n}
\rput(9.59,0.455){$L$}
\usefont{T1}{ptm}{m}{n}
\rput(8.0,0.215){$R$}
\psline[linewidth=0.02cm,linestyle=dotted,dotsep=0.16cm](6.4,0.45)(9.1,0.45)
\psline[linewidth=0.02cm,linestyle=dotted,dotsep=0.16cm](6.4,-0.75)(9.1,-0.75)
\psline[linewidth=0.02cm,linestyle=dotted,dotsep=0.16cm](6.7,-0.15)(7.3,-0.15)
\psline[linewidth=0.02cm,linestyle=dotted,dotsep=0.16cm](10.0,-0.75)(10.5,-0.75)
\psline[linewidth=0.02cm,linestyle=dotted,dotsep=0.16cm](10.0,0.45)(10.5,0.45)
\psline[linewidth=0.02cm,linestyle=dotted,dotsep=0.16cm](10.0,-0.15)(10.5,-0.15)
\end{pspicture} 
}
\caption{The transformation~$V$ in momentum space.}
\label{figVtrans}
\end{figure}
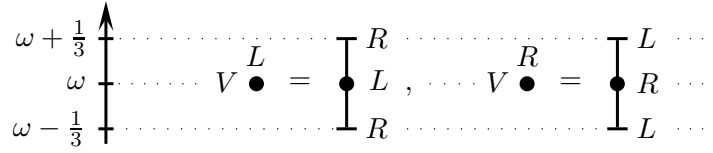

Let us first discuss the effect of the transformation~$V$. A left-handed spinor is mapped to
\[ V \begin{pmatrix} 1 \\ 0 \end{pmatrix} = 
\begin{pmatrix} 1 \\ 0 \end{pmatrix} + \frac{i}{2}\: e^{it/3}\, \begin{pmatrix} 0 \\ 1 \end{pmatrix}
+ \frac{i}{2}\: e^{-it/3}\, \begin{pmatrix} 0 \\ 1 \end{pmatrix} \:. \]
Thus two right-handed contributions are generated, whose frequency differ from the frequency
of the left-handed component by~$\pm1/3$. Similarly, a right-handed spinor is mapped to
\[ V \begin{pmatrix} 0 \\ 1 \end{pmatrix} = 
\begin{pmatrix} 0 \\ 1 \end{pmatrix} + \frac{i}{2}\: e^{it/3}\, \begin{pmatrix} 1 \\ 0 \end{pmatrix}
+ \frac{i}{2}\: e^{-it/3}\, \begin{pmatrix} 1 \\ 0 \end{pmatrix} \:, \]
generating two left-handed components with frequencies shifted by $\pm 1/3$.
Again plotting the frequencies vertically, we depict the transformation~$V$ as in Figure~\ref{figVtrans}.
The same notation is also used in Figure~\ref{figsnail} for the transformed
plane-wave solutions.

The inner product~$\Sl .| \chi_L . \Sr$ in~\eqref{SL2} only gives a contribution if the
arguments on the left and right have the opposite chirality.
Since the transformed plane-wave solutions~$V e_{k,c}$ have a fixed chirality
at every lattice point, one sees in particular that~\eqref{SL2} vanishes if~$\mu$ is chosen
as a constant. By adding to the constant~$\mu=1$ contributions with different momenta,
we can connect the different lattice points in Figure~\ref{figsnail}. This leads us to the ansatz
\beq \label{muform}
\mu(t, \varphi) = 1 + \mu_\text{hor}(t, \varphi) + \mu_\text{vert}(t, \varphi) \:,
\eeq
where the last two summands should describe the horizontal respectively vertical arrows
in Figure~\ref{figsnail}. For the horizontal arrows we can work similar to~\eqref{vertical} with a
spatially-dependent conformal transformation. However, in order to make sure that the
left-handed component generated by~$V$ (corresponding to the two $L$s at the very right
of Figure~\eqref{figsnail}) are not connected horizontally, we include two Fourier modes
which shift the frequency by~$\pm 2/3$,
\beq \label{muhor}
\mu_\text{hor}(t, \varphi) = a(\varphi) \left( 1 - e^{\frac{2 i t}{3}} - e^{-\frac{2 i t}{3}} \right) ,
\eeq
where~$a$ has the Fourier decomposition
\beq \label{aser}
a(\varphi) = \sum_{k =1}^\infty \left( a_k\, e^{i k \varphi} + \overline{a_{-k}}\, e^{-i k \varphi} \right) \:.
\eeq
For the vertical arrows we must be careful that the left-handed contribution of~$V \e_{k,L}$
is not connected to the right-handed component of~$V \e_{k,R}$, because then the arrow
would have the wrong direction. To this end, we avoid integer frequencies.
Instead, we work with the frequencies in~$\Z \pm 1/3$, because they
connect the left-handed component~$V \e_{k,R}$ to the right-handed component of~$V \e_{k,L}$.
This leads us to the ansatz
\beq \label{bser}
\mu_\text{vert}(t, \varphi) = \mu_\text{vert}(t) = \sum_{n \in \Z} e^{i n t}
\left(b_n \,e^{\frac{it}{3}} + \overline{b_{-n}}\, e^{\frac{-it}{3}} \right) .
\eeq
The ans\"atze~\eqref{aser} and~\eqref{bser} ensure that~$\mu$ is real-valued.
Moreover, by choosing the Fourier coefficients sufficiently small, one can clearly arrange that
the first summand in~\eqref{muform} dominates, so that~$\mu$ is strictly positive.
We thus obtain the following result.

\begin{Prp} Assume that the Fourier coefficients~$a_k$ and~$b_n$
in~\eqref{aser} and~\eqref{bser} are sufficiently small and
that almost all Fourier coefficients are non-zero.
Then the function~$\mu$ defined by~\eqref{muhor} and~\eqref{muform}, is strictly positive.
Consider the Dirac operator~\eqref{Dirdef} with~$U$ and~$f$ according
to~\eqref{Utdef} (for some fixed~$\nu \in \R \setminus \{0\}$) and~\eqref{lambdadef}.
Then the chiral index of the fermionic signature operator (see Definition~\ref{defind}) 
is finite and~$\ind \Sig = p$.
\end{Prp}

We finally discuss the form of the Dirac operator in position space.
Substituting~\eqref{Diracform} into~\eqref{Dirdef} and using
the above form of~$U$ and~$f$, the Dirac operator~$\tilde{\Dir}$
can be computed in closed form. Similar as discussed in the previous section,
the Dirac operator contains a nonlocal integral operator with a singular potential.
Moreover, the transformation~$U$ modifies the Dirac matrix~$\gamma^1$ to
\[ \gamma^1 \rightarrow U \gamma^1 U^* = 
\frac{1}{1+ \nu^2 \cos^2(t/3)} \Big( \big(1-\nu^2 \cos^2(t/3) \big) \:\gamma^1 - 2 \nu \cos^2(t/3)\: \gamma^2
\Big) \:, \]
where
\[ \gamma^2 := \begin{pmatrix} i & 0 \\ 0 & -i \end{pmatrix}\:. \]
Thus the representation of the Dirac matrices becomes time-dependent; this is the main effect
of the vectorial transformation~$U$. This transformation changes the first order terms in the
Dirac equation. Moreover, the conformal transformation also changes the first order terms
just as in~\eqref{Dir1} by a prefactor~$1/f$.

\section{Examples Illustrating the Homotopy Invariance} \label{secex3}
We now give two examples to illustrate our considerations on the homotopy invariance of
the chiral index.
We begin with an example which shows that the dimension of the kernel of~$\Sig_L$
does not need to be constant for deformations which are continuous in~$\Lin(\H_0)$.
It may even become infinite-dimensional.

\begin{Example} \label{exinstable}
{\em{ We consider the space-time~$\scrM=(0, T) \times S^1$ with coordinates~$t \in (0, T)$
and~$\varphi \in [0, 2 \pi)$ endowed with the Minkowski metric
\[ ds^2 = dt^2 - d\varphi^2 \:. \]
We again choose two-component complex spinors with the spin scalar product~\eqref{ssprod}.
The Dirac operator is chosen as
\[ \Dir = i \gamma^0 \partial_t + i \gamma^1 \partial_\varphi \:, \]
where the Dirac matrices are again given by~\eqref{gammadef}.
The pseudoscalar matrix and the chiral projectors are again
chosen according to~\eqref{pseudoc} and~\eqref{defchir}.

We consider the massless Dirac equation
\[ \Dir \psi = 0 \:. \]
This equation~\eqref{Dir0} can be solved by plane wave solutions, which we write as
\beq \label{eLR2}
\e_{k,L}(\zeta) = \frac{1}{2 \pi}\: e^{+i k t + i k \varphi} \begin{pmatrix} 1 \\ 0 \end{pmatrix} \:,\qquad
\e_{k,R}(\zeta) = \frac{1}{2 \pi}\: e^{-i k t + i k \varphi} \begin{pmatrix} 0 \\ 1 \end{pmatrix} \:,
\eeq
where~$k \in \Z$ (the indices~$L$ and~$R$ denote the left- and right-handed components; at the
same time they propagate to the left respectively right). By direct computation, one verifies
that~$(\e_{k,c})_{k \in \Z, c \in \{L,R\}}$ is an orthonormal basis of~$\H_0$.

We next compute the space-time inner product~\eqref{stip},
\begin{align*}
\bra \e_{k,R} | \e_{0, L} \ket &= \int_0^T dt \int_0^{2 \pi} d\varphi\; 
\Sl \e_{k,R}(t,\varphi) \,|\, \e_{0, L}(t,\varphi) \Sr
= \frac{1}{2 \pi}  \int_0^T \delta_{k,0} \,dt = \frac{T}{2 \pi}\: \delta_{k,0} \\
\bra \e_{k,R} | \e_{k', L} \ket 
&= \frac{1}{2 \pi} \: \delta_{k,k'} \int_0^T e^{2 i k t} \,dt = \frac{e^{2 i k T} - 1}{4 \pi i k}
\quad \text{($k' \neq 0$)} \\
\bra \e_{k,L} | \e_{k', L} \ket &= 0 = \bra \e_{k,R} | \e_{k', R} \ket \:.
\end{align*}
Thus the fermionic signature operator~$\Sig$ is invariant on the subspaces~$\H^{(k)}_0$
generated by the basis vectors~$\e_{k,L}$ and~$\e_{k,R}$. Moreover, in these bases it has the
matrix representations
\[ \Sig|_{\H^{(0)}_0} = \frac{T}{2 \pi} \begin{pmatrix} 0 & 1 \\ 1 & 0 \end{pmatrix} \qquad \text{and} \qquad
\Sig|_{\H^{(k)}_0} = \frac{1}{4 \pi i k}\:
\begin{pmatrix} 0 & e^{2 i k T} - 1 \\ e^{-2 i k T} - 1 & 0 \end{pmatrix} \quad (k \neq 0)\:. \]

If~$T \not \in \pi \Q$, the matrix entries~$e^{\pm 2 i k T} - 1$ are all non-zero. As a consequence,
the operators~$\Sig_L|_{\H_L}$ and~$\Sig_R|_{\H_R}$ are both injective.
Thus~$\Sig$ has finite chiral chiral index in the massless odd case (see Definition~\ref{defind0}).
If~$T \in \pi \Q$, however, the chiral index vanishes for all~$k$ for which~$2 k T$ is a multiple of~$2 \pi$.
As a consequence, the operators~$\Sig_L|_{\H_L}$ and~$\Sig_R|_{\H_R}$
both have an infinite-dimensional kernel, so that~$\Sig$ does not have a finite chiral index.
}} \QEDrem
\end{Example}
This example also explains why we need additional assumptions like those in
Theorems~\ref{thmstable} and~\ref{thmstablem0}. In particular, 
when considering homotopies of space-time or of the Dirac operator,
one must be careful to ensure that the chiral index remains finite along the chosen path.

We next want to construct examples of homotopies to which the stability result of
Theorem~\ref{thmstablem0} applies. To this end, it is convenient to work similar to~\eqref{conform}
with a conformal transformation.
\begin{Example} \label{exstable}
{\em{ As in Example~\ref{exinstable} we consider the space-time~$(0,T) \times S^1$,
but now with the conformally transformed metric
\[ d\tilde{s}^2 = f(t)^2 \left( dt^2 - d\varphi^2 \right) \]
where~$f$ is a non-negative $C^2$-function with
\[ \supp f \subset (-T,T) \qquad \text{and} \qquad f(0) > 0 \:. \]
Similar to~\eqref{vertical} and the computation thereafter, 
transforming the plane-wave solutions~\eqref{eLR2} conformally
to~$\tilde{\e}_{k, L\!/\!R} = f(t)^{-\frac{1}{2}}\: \e_{k, L\!/\!R}$,
we again obtain an orthonormal basis of~$\H_0$ and
\begin{align*}
\bra \tilde{\e}_{k,R} | \tilde{\e}_{k', L} \ket 
&= \frac{1}{2 \pi} \: \delta_{k,k'} \int_0^T f(t)\: e^{2 i k t} \,dt  \\
\bra \tilde{\e}_{k,L} | \tilde{\e}_{k', L} \ket &= 0 = \bra \tilde{\e}_{k,R} | \tilde{\e}_{k', R} \ket
\end{align*}
for all~$k, k' \in \Z$.

The integration-by-parts argument
\begin{align*}
\int_0^T f(t)\: e^{2 i k t} \,dt &=
\frac{1}{2 i k} \int_0^T f(t)\: \frac{d}{dt} e^{2 i k t} \,dt = 
-\frac{f(0)}{2 i k} - \frac{1}{2 i k} \int_0^T f'(t)\: e^{2 i k t} \,dt \\
&= -\frac{f(0)}{2 i k} -\frac{f'(0)}{4 k^2}
-\frac{1}{4 k^2} \int_0^T f''(t)\: e^{2 i k t} \,dt 
\end{align*}
shows that the space-time inner products have a simple explicit asymptotics for large~$k$ given by
\[ \bra \tilde{\e}_{k,R} | \tilde{\e}_{k', L} \ket = \frac{f(0)}{4 \pi i k}\; \delta_{k,k'}
+ \O \Big( \frac{1}{k^2} \Big) \:. \]
Hence the operator~$\Sig_L$ has the form
\[ \Sig_L \e_{k,L} = c_k\, \e_{k,R} \]
with coefficients~$c_k$ having the asymptotics
\[ c_k = \frac{f(0)}{4 \pi i k} + \O \Big( \frac{1}{k^2} \Big) \:. \]
From this asymptotics we can read off the following facts. First, it is obvious that~$\Sig_L|_{\H_L}$
has a finite-dimensional kernel. Exchanging the chirality, the same is true for~$\Sig_R|_{\H_R}$,
implying that~$S$ has a finite chiral index (according to Definition~\ref{defind0}).
Next, the vectors in the image of~$\Sig_L$ are in the Sobolev space $W^{1,2}$,
\[ \Sig_L|_{\H_L} \::\: \H_L \rightarrow \H_R \cap W^{1,2}(S^1, \C^2) \:. \]
Moreover, the image of this operator is closed (in the $W^{1,2}$-norm).
Finally, our partial integration argument also yields that
\[ \|\Sig_L \psi\|_{W^{1,2}} \leq |f|_{C^2}\: \|\psi\|_{\H_0} \:, \]
showing that the family of signature operators is norm continuous
for a $C^2$-homotopy of functions~$f$.

Having verified the assumptions of Theorem~\ref{thmstablem0},
we conclude that the chiral index in the massless odd case is invariant
under $C^2$-homotopies of the conformal function~$f$, provided that~$f(0)$
stays away from zero.
}} \QEDrem
\end{Example}

\section{Conclusion and Outlook} \label{secoutlook}
Our analysis shows that the chiral index of a fermionic signature operator
is well-defined and in general non-trivial. Moreover, it is a homotopy invariant provided
that the additional conditions stated in Theorems~\ref{thmstable} and~\ref{thmstablem0}
are satisfied. As already mentioned at the end of the introduction,
the physical and geometric meaning of this index is yet to be explored.

We now outline how our definition of the chiral index could be generalized or
extended other situations. First, our constructions also apply in the {\em{Riemannian setting}}
by working instead of causal fermion systems with so-called Riemannian fermion systems
or general {\em{topological fermion systems}} as introduced in~\cite{topology}.
In this situation, one again imposes a pseudoscalar operators~$\Gamma(x) \in \Lin(\H)$
with the properties~\eqref{pseudodef}. Then all constructions in Section~\ref{seccfs}
go through. Starting on an even-dimensional Riemannian spin manifold, one can proceed as
explained in~\cite{topology} and first construct a corresponding topological fermion system.
For this construction, one must choose a particle space, typically of eigensolutions of the Dirac equation.
Once the topological fermion system is constructed, one can again work with the index of Section~\ref{seccfs}.
If the Dirac operator anti-commutes with the pseudoscalar operator, one can choose the
particle space~$\H$ to be invariant under the action of~$\Gamma$. This gives a
decomposition of the particle space into two chiral subspaces,
$\H = \H_L \oplus \H_R$. Just as explained in Section~\ref{secodd},
this makes it possible to introduce other indices by restricting the chiral signature operators
to~$\H_L$ or~$\H_R$. Moreover, one could compose the operators from the left
with the projection operators onto the subspaces~$\H_{L\!/\!R}$ and consider the Noether indices
of the resulting operators.

Another generalization concerns space-times of {\em{infinite lifetime}}.
Using the constructions in~\cite{infinite}, in such space-times one can still introduce
the fermionic signature operator~$\Sig_m$ provided that the space-time satisfies
the so-called mass oscillation property. By inserting chiral projection operators, one
can again define chiral signature operators~$\Sig^{L\!/\!R}_m$ and define the
chiral index as their Noether index. Also the stability results of Theorems~\ref{thmstable}
and~\ref{thmstablem0} again apply. It is unknown whether the resulting indices have a
geometric meaning. Since~$\Sig_m$ depends essentially on the asymptotic form of
the Dirac solutions near infinity, the corresponding chiral indices should encode 
information on the metric and the external potential in the asymptotic ends.

We finally remark that the fermionic signature operator could be {\em{localized}}
by restricting the space-time integrals to a measurable subset~$\Omega \subset \scrM$.
For example, one can introduce a chiral signature operator~$\Sig_L(\Omega)$
similar to~\eqref{stipLR} and~\eqref{SLRdef2} by
\[ ( \phi \,|\, \Sig_L(\Omega)\, \psi) = 
\int_\Omega \Sl \psi \,|\, \chi_L \,\phi \Sr_x\: d\mu_\scrM\:. \]
Likewise, in the setting of causal fermion systems, one can modify~\eqref{sigLint} to
\[ \Sig_L(\Omega) = -\int_\Omega x \,\chi_L\: d\rho(x) \:. \]
The corresponding indices should encode information on the behavior of the Dirac solutions
in the space-time region~$\Omega$.

\Thanks {{\em{Acknowledgments:}}
I would like to thank Niky Kamran and Hermann Schulz-Baldes for helpful discussions.


\providecommand{\bysame}{\leavevmode\hbox to3em{\hrulefill}\thinspace}
\providecommand{\MR}{\relax\ifhmode\unskip\space\fi MR }
\providecommand{\MRhref}[2]{%
  \href{http://www.ams.org/mathscinet-getitem?mr=#1}{#2}
}
\providecommand{\href}[2]{#2}

\end{document}